\documentstyle[preprint,aps,epsfig]{revtex}

\begin{document}
\draft
\title{Phenomenological analysis of properties of the RH Majorana neutrino in the seesaw mechanism}
\author{Haijun Pan$^1$\thanks{Email: phj@mail.ustc.edu.cn} and 
	G. Cheng$^{2,3}$\thanks{Email: gcheng@ustc.edu.cn}}
\address{\it
$^1$Lab of Quantum Communication and Quantum Computation, 
and Center of Nonlinear Science,\\
University of Science and Technology of China, Hefei, Anhui, 230026, P.R.China\\
$^2$CCAST(World Laboratory), P.O.Box 8730, Beijing 10080, P.R.China\\
$^3$Department of Astronomy and Applied Physics, \\
University of Science and Technology of China,Hefei, Anhui, 230026, P.R.China}
\date{\today}
\maketitle

\begin{abstract}
As an extension of our previous work in the seesaw mechanism, we analyze 
the influence of $U_{e3}$ on the properties (masses and mixing) of the RH
Majorana neutrinos in three flavors. The quasidegenerate light neutrinos
case is also considered. Assuming the hierarchical Dirac neutrino masses, 
we find the heavy Majorana neutrino mass spectrum is either hierarchical 
or partial degenerate if $\theta _{23}^{\nu }$ is large. We show that
degenerate RH Majorana masses correspond to maximal RH mixing angle while
hierarchical ones correspond to the RH mixing angles which scale linearly
with the mass ratios of the Dirac neutrino masses. An interesting analogue
with the behavior of the matter-enhanced neutrino conversion is also
presented.
\end{abstract}


\newpage \baselineskip20pt

\section{Introduction}

By adding the heavy right-handed (RH) neutrinos $\nu _{R}$, the seesaw
mechanism \cite{ss}\ provides a very natural and attractive explanation 
of the smallness of the neutrino masses compared to the masses of the 
charged fermions. In the seesaw mechanism the mass matrix of the 
left-handed (LH) neutrino has the following form \cite{smirnov}: 
\begin{equation}
	m_{\nu }=m_{D}M^{-1}m_{D}^{T},  \label{seesaw}
\end{equation}
where $m_{D}$ is the Dirac mass matrix and $M$ is the Majorana mass matrix 
for the right-handed neutrino components. According to some kind of 
quark-lepton analogy suggested in Grand Unified Theories (GUTs), the structure 
of the Dirac mass matrix $m_{D}$ are similar to that in the quark sector 
\cite{smirnov}. However, the scale of the RH Majorana neutrino masses is not 
precisely known. In various theoretical model, $\nu _{R}$ can be the unification 
scale ($\sim 10^{16}~{\rm GeV}$) or the intermediate scale 
($\sim 10^{9}-10^{13}~{\rm GeV}$) \cite{falcone}. To understand possible 
unification of particles and interactions, it is crucial to know if this
scale associated with some new physics and at what energy it eventually
happens \cite{falcone}.

The seesaw mechanism can give us hints about the neutrino properties in two
aspects \cite{ss,kuo}. The first is fixing $M$ by some ansatz and
predicting masses and mixing of the light neutrinos. With the
increasing of the data from the low neutrino experiments, it becomes
pressing and practical to determine the structure of the RH Majorana
neutrinos from the light neutrino masses and mixing implied by experiments 
\cite{falcone,kuo,previous}.

The knowledge about neutrino masses and mixing comes mainly from three kinds
of neutrino oscillation experiments: the solar \cite{solar} and atmospheric 
\cite{sk} neutrino deficits, LSND reactor experiments \cite{LSND}. The 
observation of the CHOOZ Collaboration \cite{chooz} implied a small $U_{e3}$ 
\cite{ak}
\begin{equation}
	\left| U_{e3}\right| =\sin \theta _{13}^{\nu }\leq 0.13-0.23
\end{equation}
and so the atmospheric neutrino oscillation $\nu _{\mu }\leftrightarrow 
\nu _{\tau }$ decouples approximately from the solar neutrino oscillation 
$\nu _{e}\leftrightarrow \nu _{\mu }$. Analysis of the atmospheric neutrino
deficit observed by the Super-Kamiokande Collaboration yield the
mass-squared difference \cite{sk}
\begin{equation}
	\Delta m_{23}^{2}=5.9\times 10^{-3}~{\rm eV}^{2}
\end{equation}
with the almost maximal mixing $\sin ^{2}2\theta _{23}^{\nu }=1$ \cite{sk}.
In contrast, there exists two different oscillation mechanism yielding four
possible solutions \cite{bahcall}\ to the solar neutrino problem:
''Just-so'' mechanism (or VO, i.e. the long wavelength vacuum oscillations)
with 
\begin{equation}
	\left( \Delta m_{12}^{2},\sin ^{2}2\theta _{12}^{\nu }\right) 
		=\left(6.5\times 10^{-11}~{\rm eV}^{2},0.75\right)
\end{equation}
and the Mikheyev-Smirnov-Wolfenstein (MSW) mechanism \cite{msw} (the
matter-enhanced oscillation) with 
\begin{equation}
\begin{array}{clr}
	\left( \Delta m_{12}^{2},\sin ^{2}2\theta _{12}^{\nu }\right) 
	& =\left(1.8\times 10^{-5}~{\rm eV}^{2},0.76\right) & {\rm (LMA)}, \\ 
	& =\left( 7.9\times 10^{-8}~{\rm eV}^{2},0.96\right) & {\rm (LOW)}, \\ 
	& =\left( 5.4\times 10^{-6}~{\rm eV}^{2},6.0\times 10^{-3}\right) & {\rm (SMA)}.
\end{array}
\label{data}
\end{equation}
Here LMA (SMA) refers to large (small) mixing angle and LOW stands for low
mass or possibility. All the above we take are the best-fit values. 
For the convenience of late discussion, we present here the regions of the
mass-squared differences $\Delta m_{12}^{2}$ from Bahcall and $\Delta m_{23}^{2}$ 
obtained from SK: 
\begin{equation}
\begin{array}{lrll}
	\Delta m_{12}^{2}{\rm :} 
		& 4\times 10^{-12}\sim 6\times 10^{-9}~{\rm eV}^{2} &  & {\rm (VO)}, \\ 
		& 6\times 10^{-6}\sim 3\times 10^{-4}~{\rm eV}^{2} &  & {\rm (LMA)}, \\ 
		& 3\times 10^{-8}\sim 2\times 10^{-7}~{\rm eV}^{2} &  & {\rm (LOW)}, \\ 
		& 4\times 10^{-6}\sim 1\times 10^{-5}~{\rm eV}^{2} &  & {\rm (SMA)}, \\ 
	\Delta m_{23}^{2}{\rm :} 
		& 1\times 10^{-3}\sim 1\times 10^{-1}~{\rm eV}^{2} &  & {\rm (From~SK)}.
\end{array}
\end{equation}
Another neutrino oscillation experiment, LSND, indicates the mass-squared
difference 
\begin{equation}
\Delta m_{LSND}^{2}\sim 1~{\rm eV}^{2}
\end{equation}
with the mixing angle $\sin ^{2}2\theta _{LSND}^{\nu }\sim 10^{-3}-10^{-2}$.
Four neutrinos are needed to accommodate all the three mass-squared differences. 
However, the LSND results were not confirmed by the recent KARMEN experiment and 
we will just set it aside \cite{fritzsch}.

In the basis that the Dirac mass matrix of charged leptons is diagonal, the
seesaw matrix \cite{smirnov}, $S$, can be written as: 
\begin{equation}
	S=D_{L}^{T}U.
\end{equation}
Here $D_{L}$ is just $U_{0}$ in Ref.\ \cite{previous} which is the left-handed 
transformation to diagonalize $m_{D}$. We will ignore the CP-violating effect 
(i.e. all the mixing matrices entered in the seesaw mechanism are real). It is 
convenient to set $D_{L}=I$ and include its influence in the structure of $U$. 
In this paper we will consider how a nonzero $U_{e3}$ will affect the RH neutrino 
masses and mixing as an extension of our previous analysis \cite{previous}. There 
are three reason for us to devote to this topic: one is that $\theta _{13}^{\nu }$ 
can be comparable with and can even be far larger than $\theta _{12}^{\nu }$ in 
the allowed region of the small mixing MSW effect. The other is that $S$ contains 
the contribution from $D_{L}^{T}$. \ If assuming $m_{D}$ has the same 
structure, not only the mass scale but also the mixing matrix, with that in quark 
sector, one has \cite{falcone} 
\begin{equation}
	D_{L}=\left( 
		\begin{array}{ccc}
			1-\frac{1}{2}\lambda ^{2} & \lambda & \lambda ^{4} \\ 
			-\lambda & 1-\frac{1}{2}\lambda ^{2} & \lambda ^{2} \\ 
			\lambda ^{3}-\lambda ^{4} & -\lambda ^{2} & 1-\frac{1}{2}\lambda ^{2}
		\end{array}
		\right)
\end{equation}
with $\lambda \approx 0.22$. Then $S_{12}$ and $S_{13}$ have the following
values:
\begin{equation}
	S_{12}=U_{e2}-\lambda U_{\mu 2}-\frac{1}{2}\lambda ^{2}U_{e2}
		+\left( \lambda ^{3}-\lambda ^{4}\right) U_{\tau 2},
\end{equation}
\begin{equation}
	S_{13}=U_{e3}-\lambda U_{\mu 3}-\frac{1}{2}\lambda ^{2}U_{e3}
		+\left( \lambda ^{3}-\lambda ^{4}\right) U_{\tau 3}.
\end{equation}
Since $\lambda \gg U_{e2}\sim 10^{-2}$ in the SMA region, one can see that 
$S_{12}\sim S_{13}\sim -\lambda $ even when $\theta _{13}^{\nu }=0$. The
third is that, from Eq.\ (\ref{A}) and Eq.\ (\ref{X}) (see in Sec.\ \ref{II}), 
the coefficient of $U_{e3}$ in $A$ is the maximal one. Moreover, from 
Eq.\ (\ref{B}) and Eq.\ (\ref{Y}), the coefficient of $U_{\tau 1}$ in $B$ 
is also the maximal one. So the scale and structure of the Majorana neutrino 
may be sensitive, particularly in the SMA region i.e. when $\theta _{12}^{\nu }$ 
is small, to the value of $U_{e3}$ and $U_{\tau 1}$, that is, to the values of 
$\theta _{12}^{\nu }\ $ and $\theta _{13}^{\nu }$.

In section II, the quasidegenerate mass case is briefly discussed. In section 
III, the hierarchical mass case is studied in two possibilities according to 
whether $\theta _{12}^{\nu }$ is large or small. Finally, in section V we 
summary our main results.


\section{quasidegenerate spectrum}
\label{II}
In Ref.\ \cite{previous} we have obtained the three eigenvalues of $M$ 
which can be expressed as 
\begin{equation}
	M_{1}=F\overline{M}_{1},\ \ \ 
	M_{2}=F\overline{M}_{2},\ \ \ 
	M_{3}=F\overline{M}_{3},
\end{equation}
and the eigenvectors of $M$  
\begin{equation}
	V_{ij}=\frac{Y_{ij}+\overline{M}_{j}X_{ij}}
			{\left( Y_{jj}+\overline{M}_{j}X_{jj}\right) 
			+\overline{M}_{j}^{-1}-{\rm Tr}Y}V_{jj}
		\ \ \ \left( i,j=1,2,3~{\rm and}~i\neq j\right) ,
\end{equation}
where 
\begin{equation}
	F=\left( \frac{m_{1D}^{2}m_{2D}^{2}m_{3D}^{2}}{m_{1}m_{2}m_{3}}\right) 
		^{\frac{1}{3}},
\end{equation}
and $\overline{M}_{i}=e^{-2\eta _{3}\lambda _{3}-2\sqrt{3}\eta _{8}\lambda
_{8}}$ which satisfy 
\begin{equation}
	\overline{M}_{1}^{-1}+\overline{M}_{2}^{-1}+\overline{M}_{3}^{-1}
		=X_{11}+X_{22}+X_{33}\equiv A,  \label{A}
\end{equation}
\begin{equation}
	\overline{M}_{1}+\overline{M}_{2}+\overline{M}_{3}
		=Y_{11}+Y_{22}+Y_{33}\equiv B.  \label{B}
\end{equation}
Here $\lambda _{3,8}$ are diagonal Gell-Mann matrices. The elements of $X$
and $Y$ can be expressed as 
\begin{equation}
	X_{ij}=\frac{1}{\overline{m}_{iD}\overline{m}_{jD}}
		\sum_{k=1}^{3}\overline{m}_{\nu k}S_{ik}S_{jk},  
\label{X}
\end{equation}
\begin{equation}
	Y_{ij}=\overline{m}_{iD}\overline{m}_{jD}
		\sum_{k=1}^{3}\frac{1}{\overline{m}_{\nu k}}S_{ik}S_{jk},  
\label{Y}
\end{equation}
where $\overline{m}_{iD}=e^{-\xi _{3}\lambda _{3}-\sqrt{3}\xi _{8}\lambda
_{8}}$ and $\overline{m}_{\nu k}=e^{-2\kappa _{3}\lambda _{3}-2\sqrt{3}
\kappa _{8}\lambda _{8}}$. 

When $m_{0}=m_{1}\approx m_{2}\approx m_{3}$, we define $\delta _{12}$ and $\delta _{13}$ as 
\begin{equation}
	m_{2}=m_{0}\left( 1+\delta _{12}\right) ,\ \ \ 
	m_{3}=m_{0}\left( 1+\delta _{13}\right)
\end{equation}
which satisfy $\delta _{12}\ll \delta _{13}$. From the above relations it is easy for 
one to obtain 
\begin{equation}
M_{i}\approx \frac{m_{iD}^{2}}{m_{0}},\ \ \ i=1,2,3  \label{deg1}
\end{equation}
and 
\begin{equation}
	\beta _{ij}\approx 
		-\frac{m_{iD}}{m_{jD}}\left( H_{ij}\left( \delta _{12},\delta _{13}\right) 
		-gH_{ij}\left( \frac{\delta _{12}}{1+\delta _{12}}, 
				\frac{\delta _{13}}{1+\delta _{13}}\right) \right) ,
	\ \ \ 1\leq i<j\leq 3,
\end{equation}
where the function $H_{ij}$ is defined as 
\begin{equation}
	H_{ij}\left( \delta _{12},\delta _{13}\right) 
		=\delta _{12}S_{i2}S_{j2}+\delta _{13}S_{i3}S_{j3}
\end{equation}
and 
\begin{equation}
	g=\left\{ 
	\begin{array}{ccl}
		\frac{m_{2D}^{2}}{m_{3D}^{2}}, 
		&  & {\rm for~}\beta _{12}{\rm ~and~}\beta _{13} \\ 
		\frac{m_{1D}^{2}}{m_{3D}^{2}}, 
		&  & {\rm for~}\beta _{23}
\end{array}
\right. .
\end{equation}

We can see the RH mixing angles are small. However their scales are sensitive 
to the inputs of $S_{12}$, $S_{13}$, $\delta _{12}$ and $\delta _{13}$. For 
example, $\beta _{13}\approx \frac{m_{1D}}{m_{3D}}\delta _{12}S_{12}S_{32}$ 
when $S_{13}=0$ and $\beta _{13}\longrightarrow 0$ when $\frac{S_{13}}{S_{32}}
\approx \frac{\delta _{12}}{\delta _{13}}$. Despite this sensitivity, we find 
\begin{equation}
	\beta _{12}\approx -\frac{m_{1D}}{m_{2D}}\delta _{12}S_{12}S_{22}, \ \ \
	\beta _{13}\approx \frac{m_{1D}}{m_{3D}}\delta _{12}S_{12}S_{32}, \ \ \
	\beta _{23}\approx \frac{m_{2D}}{m_{3D}}\delta _{13}S_{23}S_{33}
\end{equation}
when $S_{13}=0$ and for SMA 
\begin{equation}
	\beta _{12}\approx -\frac{m_{1D}}{m_{2D}}\delta _{13}S_{13}S_{23}, \ \ \
	\beta _{13}\approx \frac{m_{1D}}{m_{3D}}\delta _{13}S_{13}S_{33}, \ \ \
	\beta _{23}\approx \frac{m_{2D}}{m_{3D}}\delta _{13}S_{23}S_{33}
\end{equation}
when $S_{13}\gtrsim S_{12}$.

If assuming $m_{0}\sim 1 {\rm eV}$, one has $M_{1}=1.6\times 10^{3}~{\rm GeV}$, $M_{2}=1.8\times 10^{8}~{\rm GeV}$ and $M_{3}=1.2\times 10^{13}~{\rm GeV}$. Note 
that $M_{2,3}$ are in the intermediate scale while $M_{1}$, so suprise, is in the 
electric-weak scale. 

Assuming $D_{L}=I$, all the corresponding relations can be obtained by
replace $S$ with $U$.


\section{hierarchical spectrum}

In this case the two heavier neutrino masses can be written as $m_{2}\approx 
\sqrt{\Delta m_{21}^{2}}$ and $m_{3}\approx \sqrt{\Delta m_{31}^{2}}$. Setting 
$\theta _{23}^{\nu }=\frac{\pi }{4}$, the neutrino mixing matrix $U$ have the 
form 
\begin{equation}
	U=\left( 
	\begin{array}{ccc}
		\cos \theta _{12}^{\nu }\cos \theta _{13}^{\nu } 
		& \sin \theta _{12}^{\nu }\cos \theta _{13}^{\nu } & \sin \theta _{13}^{\nu } 
		\\ 
		-\frac{\sin \theta _{12}^{\nu }
			+\cos \theta _{12}^{\nu }\sin \theta _{13}^{\nu }}{\sqrt{2}} 
		& \frac{\cos \theta _{12}^{\nu }
			-\sin \theta _{12}^{\nu }\sin \theta _{13}^{\nu }}{\sqrt{2}} 
		& \frac{\cos \theta _{13}^{\nu }}{\sqrt{2}} 
		\\ 
		\frac{\sin \theta _{12}^{\nu }
			-\cos \theta _{12}^{\nu }\sin \theta _{13}^{\nu }}{\sqrt{2}} 
		& -\frac{\cos \theta _{12}^{\nu }
			+\sin \theta _{12}^{\nu }\sin \theta _{13}^{\nu }}{\sqrt{2}} 
		& \frac{\cos \theta _{13}^{\nu }}{\sqrt{2}}
	\end{array}
	\right) .\allowbreak
\end{equation}
In follows, we first consider the small mixing solution to the solar neutrino 
problem and then VO, LMA and LOW are embodied in a unitized framework, i.e. 
the large $\theta _{12}^{\nu }$.


\subsection{small $\protect\theta _{12}^{\protect\nu }$ (SMA)}

We will always assume $\sin ^{2}2\theta _{12}^{\nu }\gtrsim 10^{-3}$ so that
it remain in the SMA region. When $\theta _{12}^{\nu }$ is small there are
three possibilities by comparing $\theta _{13}^{\nu }$ with $\theta _{12}^{\nu }$.


\subsubsection{$\protect\theta _{12}^{\protect\nu }\gg \protect\theta _{13}
^{\protect\nu }$}

We can know from Ref.\ \cite{previous} that in this case
\begin{equation}
	M_{1}\approx f_{1}\frac{m_{1D}^{2}}{m_{2}}
			\frac{1}{\sin ^{2}\theta _{12}^{\nu }},\ \ \ 
	M_{2}\approx 2\frac{m_{2D}^{2}}{m_{3}},\ \ \
	M_{3}\approx \frac{1}{2}f_{1}^{-1}
			\frac{m_{3D}^{2}}{m_{1}}\sin ^{2}\theta _{12}^{\nu }
\end{equation}
and 
\begin{equation}
	\beta _{12}\approx -\frac{1}{\sqrt{2}}f_{1}
				\frac{m_{1D}}{m_{2D}}\cot \theta _{12}^{\nu },\ \ \ 
	\beta _{13}\approx \sqrt{2}f_{1}
				\frac{m_{1D}}{m_{3D}}\cot \theta _{12}^{\nu },\ \ \ 
	\beta _{23}\approx -\frac{m_{2D}}{m_{3D}}
\end{equation}
where $f=\frac{r_{21}}{r_{21}+\cot ^{2}\theta _{12}}$\ and $r_{21}
=\frac{m_{2}}{m_{1}}\gg 1$. The relations are obtained when 
$\theta _{13}^{\nu }\rightarrow 0$ so that 
\begin{equation}
	U=\left( 
	\begin{array}{ccc}
		\cos \theta _{12}^{\nu } & \sin \theta _{12}^{\nu } & 0 
		\\ 
		-\frac{\sin \theta _{12}^{\nu }}{\sqrt{2}} 
		& \frac{\cos \theta _{12}^{\nu }}{\sqrt{2}} & \frac{1}{\sqrt{2}} 
		\\ 
		\frac{\sin \theta _{12}^{\nu }}{\sqrt{2}} 
		& -\frac{\cos \theta _{12}^{\nu }}{\sqrt{2}} & \frac{1}{\sqrt{2}}
	\end{array}
	\right) .\allowbreak
\end{equation}
Note that, unlike what one would expect when no mixing occurs $M_{i}\propto
m_{i}^{-1}$, $M_{i}$ have rotate dependence on $m_{i}^{-1}$ in the sense: 
$M_{i}\propto m_{i_{1}}^{-1}$ where we define 
\begin{equation}
	i_{k}\equiv (i+k)\ {\rm mod}\ 3\ \ \ i,k=1,2,3.
\end{equation}
As noted in Ref.\ \cite{previous} the RH mixing angles scale linearly with
the ratios of the Dirac neutrino masses 
\begin{equation}
	\beta _{ij}\sim \frac{m_{iD}}{m_{jD}},\ \ \ 1\leq i<j\leq 3.
\end{equation}
which is different with the LH quark mixing angles where one obtains 
$\tan \theta ^{{\rm quark}}\approx \sqrt{\frac{m_{d}}{m_{s}}}$ in two-generation
case \cite{quark}.


\subsubsection{$\protect\theta _{12}^{\protect\nu }\ll \protect\theta _{13}
^{\protect\nu }$}

In this case 
\begin{equation}
	U\approx \left( 
	\begin{array}{ccc}
		1 & \sin \theta _{12}^{\nu } & \sin \theta _{13}^{\nu } 
		\\ 
		-\frac{\sin \theta _{13}^{\nu }}{\sqrt{2}} 
		& -\frac{1}{\sqrt{2}} & \frac{1}{\sqrt{2}} 
		\\ 
		-\frac{\sin \theta _{13}^{\nu }}{\sqrt{2}} 
		& -\frac{1}{\sqrt{2}} & \frac{1}{\sqrt{2}}
	\end{array}
	\right) .\allowbreak
\end{equation}
Here $U$ is not a strict orthogonal matrix as required. However it brings no
trouble in since only the relative magnitudes of the elements of $U$ are
used in our analysis. We have 
\begin{equation}
	M_{1}\approx \frac{m_{1D}^{2}}{m_{3}}
			\frac{1}{\sin ^{2}\theta _{13}^{\nu }},\ \ \ 
	M_{2}\approx 2\frac{m_{2D}^{2}}{m_{1}}\frac{\sin ^{2}\theta _{13}^{\nu }}
			{r_{21}\sin ^{2}\theta _{13}^{\nu }+1},\ \ \ 
	M_{3}\approx \frac{1}{2}\frac{m_{3D}^{2}}{m_{2}}
			\left( r_{21}\sin ^{2}\theta _{13}^{\nu }+1\right)  
	\label{m}
\end{equation}
and 
\begin{equation}
	\beta _{12}\approx -\frac{m_{1D}}{m_{2D}}
				\frac{1}{\sqrt{2}\sin \theta _{13}^{\nu }},\ \ \ 
	\beta _{13}\approx \sqrt{2}\frac{m_{1D}}{m_{3D}}\sin \theta _{13}^{\nu }
				\frac{r_{21}}{r_{21}\sin ^{2}\theta _{13}^{\nu }+1},\ \ \
	\beta _{23}\approx \frac{m_{2D}}{m_{3D}}.
\end{equation}
From Eq. (\ref{m}) one has $M_{i}\propto m_{i_{2}}^{-1}$ when $r_{21}\lesssim 
\frac{U_{\tau 2}^{2}}{U_{\tau 1}^{2}}\approx \sin ^{-1}\theta _{13}^{\nu }$ 
and $M_{1(3)}\propto m_{3(1)}^{-1}$ and $M_{2}\propto m_{2}^{-1}$ when 
$r_{21}\gg \sin ^{-1}\theta _{13}^{\nu }$.


\subsubsection{$\protect\theta _{12}^{\protect\nu }\sim \protect\theta 
_{13}^{\protect\nu }$}

For convenient, we set $U_{\tau 1}=0$, that is $\sin \theta _{13}^{\nu }
=\tan \theta _{12}^{\nu }$. So that 
\begin{equation}
	U=\left( 
	\begin{array}{ccc}
		\sqrt{\cos 2\theta _{12}^{\nu }} 
		& \sqrt{\cos 2\theta _{12}^{\nu }}\tan \theta _{12}^{\nu } 
		& \tan \theta _{12}^{\nu } 
		\\ 
		-\sqrt{2}\sin \theta _{12}^{\nu } 
		& \frac{\cos 2\theta _{12}^{\nu }}{\sqrt{2}\cos \theta _{12}^{\nu }} 
		& \frac{\sqrt{\cos 2\theta _{12}^{\nu }}}{\sqrt{2}\cos \theta _{12}^{\nu }} 
		\\ 
		0 & -\frac{1}{\sqrt{2}\cos \theta _{12}^{\nu }} 
		& \frac{\sqrt{\cos 2\theta _{12}^{\nu }}}{\sqrt{2}\cos \theta _{12}^{\nu }}
	\end{array}
	\right). \allowbreak
\end{equation}
We find 
\begin{mathletters}
\label{sma3M}
\begin{eqnarray}
	M_{1} &\approx &\frac{m_{1D}^{2}}{m_{3}}\cot ^{2}\theta _{12}^{\nu }, \\
	M_{2} &\approx &\left\{ 
		\begin{array}{ccc}
			2\frac{m_{2D}^{2}}{m_{1}}\sin ^{2}\theta _{12}^{\nu }, 
			&  & {\rm if\quad }r_{21}<r_{21}^{{\rm res}} \\ 
			\frac{1}{2}\frac{m_{3D}^{2}}{m_{2}}, 
			&  & {\rm if\quad }r_{21}>r_{21}^{{\rm res}}
		\end{array}
		\right. \\
	M_{3} &\approx &\left\{ 
			\begin{array}{ccc}
			\frac{1}{2}\frac{m_{3D}^{2}}{m_{2}}, 
			&  & {\rm if\quad }r_{21}<r_{21}^{{\rm res}} \\ 
			2\frac{m_{2D}^{2}}{m_{1}}\sin ^{2}\theta _{12}^{\nu }, 
			&  & {\rm if\quad }r_{21}>r_{21}^{{\rm res}}
		\end{array}
		\right.
\end{eqnarray}
where $r_{21}^{{\rm res}}=\frac{1}{4}\frac{m_{3D}^{2}}{m_{2D}^{2}}
\csc ^{2}\theta _{12}^{\nu }$. We have two degenerate masses $M_{2}=M_{3}$ when 
$r_{21}=r_{21}^{{\rm res}}$.

The second RH mixing angle $\beta _{23}$ is given in 
\end{mathletters}
\begin{equation}
	\tan 2\beta _{23}\approx 
		\frac{\frac{1}{2}\frac{m_{3D}^{2}}{m_{2D}^{2}}\csc ^{2}\theta _{12}^{\nu }} 		{e^{4\kappa _{3}}-\frac{1}{4}\frac{m_{3D}^{2}} {m_{2D}^{2}}
		\csc ^{2}\theta _{12}^{\nu }} 
		\approx -\frac{2m_{2D}/m_{3D}}{1-r_{21}/r_{21}^{{\rm res}}} \allowbreak
\end{equation}
or 
\begin{equation}
	\sin 2\beta _{23}\approx 
		-\frac{2m_{2D}/m_{3D}}
		{\sqrt{\left( 1-r_{21}/r_{21}^{{\rm res}}\right) ^{2}+4m_{2D}^{2}/m_{3D}^{2}}},
\end{equation}
where $\kappa _{3}=\frac{1}{4}\ln \frac{m_{2}}{m_{2}}$ \cite{previous}. The
other two RH mixing angles are both small and can be expressed in $\beta
_{23}\allowbreak $ as follows
\begin{equation}
	\beta _{12}\approx \frac{1}{\sqrt{2}\sin \theta _{12}^{\nu }} 
				\left( -\frac{m_{1D}}{m_{2D}}\cos \beta _{23} 
				+\frac{m_{1D}}{m_{3D}}\sin \beta _{23}\allowbreak \right),
\end{equation}
\begin{equation}
	\beta _{13}\approx \frac{1}{\sqrt{2}\sin \theta _{12}^{\nu }} 
				\left( -\frac{m_{1D}}{m_{2D}}\cos \beta _{23} 				+\frac{m_{1D}}{m_{3D}}\sin \beta _{23}\right).
\end{equation}

The behaviors of $M_{i}$, $\eta _{3}\left( =\frac{1}{4}\ln \frac{M_{2}}{M_{1}
}\right) $, $\eta _{8}\left( =\frac{1}{12}\ln \frac{M_{3}^{2}}{M_{1}M_{2}}
\right) $ and $\beta _{ij}$ as functions of $\kappa _{3}$ are shown in Fig.\ 
\ref{fig1}. In Fig.\ \ref{fig2} we have plotted $M_{2}$, $M_{3}$ and $\sin
^{2}2\beta _{23}$\ near $\kappa _{3}^{{\rm res}}$ where $\kappa _{3}^{{\rm 
res}}$ is the location of the resonance defined as
\begin{equation}
	e^{4\kappa _{3}^{{\rm res}}}
		\equiv \frac{1}{4}\frac{m_{3D}^{2}}{m_{2D}^{2}}\csc ^{2}\theta _{12}^{\nu }.
\end{equation}
The behavior of $\sin ^{2}2\beta _{23}$ as a function of $\kappa _{3}$ is
clearly that of a resonance peaked at $\kappa _{3}=\kappa _{3}^{{\rm res}}$,
when $\sin ^{2}2\beta _{23}=1$. We can define the resonance width $\delta
_{\kappa _{3}}$ as that of $\kappa _{3}$ around $\kappa _{3}^{{\rm res}}$
for which $\sin ^{2}2\beta _{23}$ becomes $\frac{1}{2}$ instead of the
maximum value, unity. It is given by
\begin{equation}
	\delta _{\kappa _{3}}\approx \frac{m_{2D}}{m_{3D}}.
\end{equation}
The situation is very like that in the matter-enhanced $\nu
_{e}\leftrightarrow \nu _{\mu }$ oscillation in the sun while here $r_{21}$
plays a part of the effective potential $V=2\sqrt{2}G_{f}N_{e}E_{\nu }$.
Here $G_{f}$\ is the Fermi constant, $N_{e}$ is the electron number density
of the matter and $E_{\nu }$ is the neutrino energy. When $\sin ^{2}2\beta
_{23}=1$ (that is when $\kappa _{3}^{{\rm res}}\approx 4.\allowbreak 2$ i.e. 
$m_{1}\approx 1.2\times 10^{-10}~{\rm eV}$), substituting the SMA data in 
Eq.\ (\ref{data}) into Eq.\ (\ref{sma3M}) we have 
\begin{equation}
	M_{1}\approx 1.4\times 10^{7}~{\rm GeV},\ \ \ 
	M_{2}\approx M_{3}\approx 5\times 10^{15}~{\rm GeV}.
\end{equation}


\subsection{large $\protect\theta _{12}^{\protect\nu }$}

In this case it is no need to consider the relative magnitude of $\theta
_{12}^{\nu }$ and $\theta _{13}^{\nu }$ since one always has $U_{\tau
1}\approx \frac{\sin \theta _{12}^{\nu }}{\sqrt{2}}$. We shall therefore
consider this problem in the following two possibilities according to the
magnitude of $\theta _{13}^{\nu }$.


\subsubsection{$\protect\theta _{13}^{\protect\nu }$ is tiny}

We find, when 
\begin{equation}
	\sin ^{2}\theta _{13}^{\nu }\ll 
		\min \left( \frac{m_{2}}{m_{3}}U_{12}^{2},
		\ \ \	\frac{m_{1D}^{2}}{m_{2D}^{2}}U_{\tau 3}^{2}\right) 
		\approx 
		\frac{1}{2}\frac{m_{1D}^{2}}{m_{2D}^{2}},
\end{equation}
the RH Majorana masses are 
\begin{mathletters}
\begin{eqnarray}
	M_{1} &\approx &\left\{ 
		\begin{array}{ccc}
		\frac{m_{1D}^{2}}{m_{2}}\frac{1}{\sin ^{2}\theta _{12}^{\nu }} &  		
		& \frac{m_{3}}{m_{2}}<2\frac{m_{2D}^{2}}{m_{1D}^{2}}\sin ^{2}\theta _{12}^{\nu } 
		\\ 
		2\frac{m_{2D}^{2}}{m_{3}} &  
		& \frac{m_{3}}{m_{2}}>2\frac{m_{2D}^{2}}{m_{1D}^{2}}\sin ^{2}\theta _{12}^{\nu }
		\end{array}
		\right. \\
	M_{2} &\approx &\allowbreak \left\{ 
		\begin{array}{ccc}
		2\frac{m_{2D}^{2}}{m_{3}} &  
		& \frac{m_{3}}{m_{2}}<2\frac{m_{2D}^{2}}{m_{1D}^{2}}\sin ^{2}\theta _{12}^{\nu } 
		\\ 
		\frac{m_{1D}^{2}}{m_{2}}\frac{1}{\sin ^{2}\theta _{12}^{\nu }} &  
		& \frac{m_{3}}{m_{2}}>2\frac{m_{2D}^{2}}{m_{1D}^{2}}\sin ^{2}\theta _{12}^{\nu }
		\end{array}
		\right. \\
	M_{3} &\approx &\frac{1}{2}\frac{m_{3D}^{2}}{m_{1}}\sin ^{2}\theta _{12}^{\nu }
\end{eqnarray}
and the RH mixing angles 
\end{mathletters}
\begin{equation}
	\beta _{13}\approx \sqrt{2}\frac{m_{1D}}{m_{3D}}\cot \theta _{12}^{\nu },\ \ \ 
	\beta _{23}\approx -\frac{m_{2D}}{m_{3D}}.
\end{equation}
We also give a numerical result for $\beta _{12}$ together with $M_{i}$, $
\eta _{3,8}$ and the other two mixing angles as functions of $\kappa
_{8}\left( =\frac{1}{12}\ln \frac{m_{3}^{2}}{m_{1}m_{2}}\right) $ in Fig.\ 
\ref{fig3} taking $m_{3}^{2}=0.1~{\rm eV}^{2}$ and $\kappa _{3}=2$. We also
plotted $M_{1}$, $M_{2}$ and $\sin ^{2}2\beta _{12}$ near $r_{32}^{{\rm res}
}\approx 2\frac{m_{2D}^{2}}{m_{1D}^{2}}\sin ^{2}\theta _{12}^{\nu }$ in
Fig.\ \ref{fig4}.


\subsubsection{$\protect\theta _{13}^{\protect\nu }$ is not so small}

We find, when $\theta _{13}^{\nu }$ is small and satisfies
\begin{equation}
	\sin ^{2}\theta _{13}^{\nu }\gtrsim 
		\max \left( \frac{m_{2}}{m_{3}}U_{12}^{2},
		\ \ \ \frac{m_{1D}^{2}}{m_{2D}^{2}}U_{\tau 3}^{2}\right) 
		\approx
		\frac{m_{2}}{m_{3}}\sin ^{2}\theta _{12}^{\nu },
\end{equation}
the RH Majorana neutrino masses are hierarchical 
\begin{equation}
	M_{1}\approx f_{1}^{-1}\frac{m_{D}^{2}}{m_{2}},\ \ \ 
	M_{2}\approx f_{1}\frac{m_{2D}^{2}}{m_{3}}\frac{1}{U_{\tau 1}^{2}},\ \ \ 
	M_{3}\approx \frac{m_{3D}^{2}}{m_{1}}U_{\tau 1}^{2},
\end{equation}
where $f_{1}=\left( U_{e2}^{2}+\frac{m_{3}}{m_{2}}U_{e3}^{2}+\frac{
m_{1D}^{2}m_{3}}{m_{2D}^{2}m_{2}}U_{\mu 3}^{2}\right) $ and all the three RH
mixing angles are small 
\begin{equation}
	\beta _{12}\approx -\frac{U_{e2}U_{\mu 2}+\frac{m_{3}\allowbreak }{m_{2}}
		U_{e3}U_{\mu 3}}{S_{e2}^{2}
		+\frac{m_{3}}{m_{2}}S_{\tau 3}^{2}}\frac{m_{1D}}{m_{2D}},\ \ \ 
	\beta _{13}\approx \frac{m_{1D}}{m_{3D}}\frac{U_{e1}}{U_{\tau 1}},\ \ \ 
	\beta _{23}\approx \frac{m_{2D}}{m_{3D}}\frac{U_{\mu 1}}{U_{\tau 1}}.
\end{equation}
Here we also have rotate dependence of $M_{i}$ on $m_{i}^{-1}$ and the RH
mixing angles $\beta _{ij}$ scale linearly with the ratios of the Dirac
neutrino masses.


\section{summary and discussion}

Separating the solution regions of the solar neutrino problem in two cases
according two the value of $\theta _{12}^{\nu }$, we have derived simple
relations between parameters of the RH and LH Majorana neutrino masses and
mixing in the context of the seesaw mechanism and quark-lepton symmetry
within the framework of three families. Especially, as an extension of our
previous work, we have embodied quasidegenerate light neutrino mass case and
the influence of nonzero $U_{e3}$ on the properties (masses and mixing) of
the RH Majorana neutrinos. The CP-violating effect has not included. We find

\begin{enumerate}
\item quasidegenerate neutrino spectrum leads to hierarchical RH Majorana
masses and small RH mixing angles which scale linearly with the ratios of
the Dirac masses 
\begin{equation}
	\beta _{ij}\sim \frac{m_{iD}}{m_{jD}},\ \ \ 1\leq i<j\leq 3.
\end{equation}

\item For SMA, resonance like behavior of $\sin ^{2}2\beta _{23}$ is found
when $U_{\tau 1}\approx 0$. We find when $m_{1}\approx 1.2\times 10^{-10}~
{\rm eV}$ one has $\sin ^{2}2\beta _{23}=1$, $M_{1}\approx 1.4\times 10^{7}~
{\rm GeV}$ and $M_{2}\approx M_{3}\approx 5\times 10^{15}~{\rm GeV}$. $M_{2} 
$, $M_{3}$ are near the scale of GUT and $M_{3}\approx 5\times 10^{15}~{\rm 
GeV}$ for a wide range of $r_{21}$, quantitatively, for $r_{21}$ below $
r_{21}^{{\rm res}}$.

\item The behavior $\sin ^{2}2\beta _{12}$ as a functions of $\kappa _{8}$\
if $\frac{m_{2}}{m_{1}}$ is given) is a resonance peaked at $r_{32}=r_{32}^{
{\rm res}}$ for large $\theta _{12}^{\nu }$ case while $\theta _{13}^{\nu }$
is tiny and one has two lighter degenerate RH Majorana masses at this point.
In Ref.\ \cite{previous} we have not discuss this case considering that $
m_{2}$ will far less than the lower bound of the solar neutrino solutions if
taking $m_{3}^{2}\sim 10^{-3}~{\rm eV}^{2}$ which is around the best-fit
value. However, the degenerate $M_{1}$ and $M_{2}$ could be coincident with
the experimental results considering that $m_{3}^{2}$ can reach to about $
10^{-1}~{\rm eV}^{2}$ and then one has $m_{2}^{2}\sim 10^{-11}~{\rm eV}^{2}$
which lies in the region of the vacuum explanation to the solar neutrino
anomaly but still far less than the lower bound of the LMA and LOW
solutions. For LMA and LOW, one always has $\frac{m_{3}}{m_{2}}<2\frac{
m_{2D}^{2}}{m_{1D}^{2}}\sin ^{2}\theta _{12}^{\nu }$ and so 
\begin{equation}
	M_{1}\approx \frac{m_{1D}^{2}}{m_{2}}\frac{1}{\sin ^{2}\theta _{12}^{\nu }},
	\ \ \ 
	M_{2}\approx \allowbreak 2\frac{m_{2D}^{2}}{m_{3}},
	\ \ \ 
	M_{3}\approx \frac{1}{2}\frac{m_{3D}^{2}}{m_{1}}\sin ^{2}\theta _{12}^{\nu };
\end{equation}
\begin{equation}
	\beta _{12}\approx -\frac{1}{\sqrt{2}}\frac{m_{1D}}{m_{2D}}\cot \theta _{12}^{\nu },
	\ \ \ 
	\beta _{13}\approx \sqrt{2}\frac{m_{1D}}{m_{3D}}\cot \theta _{12}^{\nu },
	\ \ \ 
	\beta _{23}\approx -\frac{m_{2D}}{m_{3D}},
\end{equation}
that is, $M_{i}\propto m_{i_{1}}^{-1}$ and $\beta _{ij}\sim \frac{m_{iD}}{
m_{jD}}\left( 1\leq i<j\leq 3\right) $. For VO, the corresponding RH masses
when $r_{32}=r_{32}^{{\rm res}}$ are $M_{1}\approx M_{2}\approx 1.2\times
10^{9}~{\rm GeV}$ and $M_{3}\approx 1.8\times 10^{17}r_{21}~
{\rm GeV}$.

\item An interesting analogue of the RH parameters around the resonance with
the behavior of the matter-enhanced neutrino conversion is presented.

\item As an accessary consequence, we present a numerical method for
calculating the physical parameters in the seesaw mechanism which usually
involving extreme large and extreme small quantities simultaneously. In our
numerical results, eigenvalues larger than unit and the corresponding
eigenvectors of $X$ are directly obtained from $X$. By solving the
eigenequation of $Y$, we obtain the inverse of the eigenvalues of $X$ larger
than unit and the corresponding eigenvectors. The validness of this method
can be verified simply by the condition that the product of the three
eigenvalues of $X$ is unit. We find this condition is satisfied well.
\end{enumerate}

The results are dependent on the precise determination of $U_{e3}$. Such a
goal is expected to be reached in the neutrino long baseline experiment,
registration of the neutrino bursts from the Galactic supernova by existing
detectors SK and SNO, and the neutrino factories \cite{goal}. In this paper
we have not discuss the case when $M_{R}$ and $m_{D}$ are complex. We hope
return to it in future.

\acknowledgements

The authors would like to express our sincere thanks to Professor T.K. Kuo
for pointing out this problem to G. Cheng during his visit at Purdue
University from January to April, 2000 and provoke our interesting in it. We
are also indebted to him for his warmly help in the research progressing and
kindness by giving us his papers being prepared. We are grateful to Dr. Du
Taijiao and Dr. Tu Tao for useful discussions. The authors are also grateful
to the convenience provided by Chinese High Performance Computing Center at
Hefei. G. Cheng is supported in part by the National Science Foundation in
China grant no.19875047.




\newpage 
\begin{figure}[tbp]
\caption{The behavior of the RH Majorana masses, mass ratios and the RH
mixing angles as functions of $\protect\kappa _{3}$ for the SMA solution to
the solar neutrino anmoly. when $U_{\protect\tau 1}=0$. We take $
m_{2}^{2}=5.4\times 10^{-6}~{\rm eV}^{2}$, $m_{3}^{2}=5.9\times 10^{-3}~{\rm 
eV}^{2}$, $\sin ^{2}2\protect\theta _{12}^{\protect\nu }=6.0\times 10^{-3}$, 
$\protect\theta _{13}^{\protect\nu }=0.0$ and $\sin ^{2}2\protect\theta 
_{23}^{\protect\nu }=1.0$.}
\label{fig1}
\end{figure}

\begin{figure}[tbp]
\caption{The behavior of $M_{2}$, $M_{3}$ and $\sin ^{2}2\protect\beta _{23}$
near $\protect\kappa _{3}^{{\rm res}}$ for the SMA solution. The values of
the parameters are the same as in Fig.\ \ref{fig1}.}
\label{fig2}
\end{figure}

\begin{figure}[tbp]
\caption{The behavior of the RH Majorana masses, mass ratios and the RH
mixing angles as functions of $\protect\kappa _{8}$ for the vacuum
oscillation solution to the solar neutrino anmoly when $U_{e3}=0$. We take $
m_{2}^{2}=6.5\times 10^{-11}~{\rm eV}^{2}$, $m_{3}^{2}=0.1~{\rm eV}^{2}$, $
\sin ^{2}2\protect\theta _{12}^{\protect\nu }=0.75$ and $\sin ^{2}2\protect
\theta _{23}^{\protect\nu }=1.0$.}
\label{fig3}
\end{figure}

\begin{figure}[tbp]
\caption{The behavior of $M_{1}$, $M_{2}$ and $\sin ^{2}2\protect\beta _{12}$
near $\protect\kappa _{8}^{{\rm res}}$ for the vacuum oscillation solution
when $U_{e3}=0$. The values of the parameters are the same as in Fig.\ \ref
{fig3}.}
\label{fig4}
\end{figure}


\begin{references}
\bibitem{ss} M. Gell-Mann, P. Ramond and R. Slansky, in: {\it Supergravity},
P. van Nieuwenhuizen and D.Z. Freedman (eds.), North Holland Publ. Co., 1979;
\newline
T. Yanagida, in Proceedings of {\em Workshop on Unified Theory and Baryon
number in the Universe}, editors O. Sawada and A. Sugamoto (KEK 1979);
\newline
R. N. Mohapatra and G. Senjanovic, Phys. Rev. Lett. {\bf 44}, 912 (1980).

\bibitem{smirnov} A. Yu. Smirnov, Phys. Rev. D {\bf 48}, 3264 (1993); 
\newline
Nucl. Phys. B. {\bf 466}, 25 (1996).

\bibitem{falcone} D. Falcone, Phys. Rev. {\bf D 61}, 097302 (2000).

\bibitem{kuo} T.K. Kuo, Guo-Hong Wu and Sadek W. Mansour, Phys. Rev. {\bf D
61}, 111301 (2000).

\bibitem{previous} Haijun Pan and G. Cheng, hep-ph/0103147.

\bibitem{solar} R. Davis, Jr., D. S. Harmer, and K. C. Hoffman, 
	Phys. Rev.	Lett. {\bf 20}, 1205 (1968); \newline
GALLEX Collaboration, P. Anselmann et al., 
	Phys. Lett. {\bf B 342}, 440 (1995); \newline
GALLEX Collaboration, W. Hampel {\sl et al.}, 
	{\sl ibid. } {\bf 388}, 364 (1996); \newline
KAMIOKANDE Collaboration, Y. Fukuda {\sl et al.}, 
	Phys. Rev. Lett. {\bf 77}, 1683 (1996); \newline
SuperKamiokande Collaboration, Y. Fukuda {\sl et al.}, hep-ex/0103032, hep-ex/0103033.

\bibitem{sk} Super-Kamiokande Collaboration, Y. Fukuda {\sl et al.}, 
	Phys. Rev. Lett. {\bf 81}, 1158 (1998); \newline
	{\sl ibid.} {\bf 82}, 1810 (1999);

\bibitem{LSND} LSND Collaboration, C. Athanassopoulos, {\sl et al.}, 
	Phys. Rev. Lett. {\bf 75}, 2650 (1995).

\bibitem{chooz} Chooz Collaboration, M.~Apollonio {\sl et al.},
	Phys.~Lett.~B~{\bf 420}, 397 (1998).

\bibitem{ak} F. Boehm {\sl et al.}, hep-ex/0003022; \newline
E. Kh. Akhmedov, G. C. Branco, and M. N. Rebelo, 
	Phys. Rev. Lett. {\bf 84}, 3535 (2000).

\bibitem{bahcall} J. Bahcall, P. Krastev, and A. Y. Smirnov, 
	Phys.Rev. D{\bf 58}, 096016 (1998); \newline
	{\sl ibid.} D{\bf 60}, 093001 (1999).

\bibitem{msw} L. Wolfenstein, Phys. Rev. {\bf D 17}, 2369 (1978); \newline
S.P. Mikheyev and A.Yu. Smirnov, Nuovo Cimento {\bf C 9}, 17 (1986).

\bibitem{fritzsch} H. Fritzsch and Zhi-zhong Xing, hep-ph/9912358.

\bibitem{quark} H. Fusaoka and Y. Koide, Phys. Rev. {\bf D 57}, 3986 (1998).

\bibitem{goal} A.Yu. Smirnov, hep-ph/0010097.
\end{references}
\end{document}